\def\meV{{\,\textrm{meV}}}
\def\fs{{\,\textrm{fs}}}
\def\ps{{\,\textrm{ps}}}
\def\K{{\,\textrm{K}}}
\def\aa{{\,\textrm{\AA}}}
\def\kjmol{{\,\textrm{kJ/mol}}}
\pdfoutput=1

\documentclass[onecolumn,superscriptaddress,preprintnumbers,amsmath,amssymb,prb,11pt]{revtex4}

\usepackage{graphicx}
\usepackage{setspace}

\usepackage{subfigure}
\usepackage{dcolumn}
\usepackage{bm}
\usepackage{color}
\newcommand{\Keywords}[1]{\par\noindent {\small{\em Keywords\/}: #1}}

\begin{document}

\title{ \bf{First-Principles Studies of Photoinduced Charge Transfer \\ in Noncovalently Functionalized Carbon Nanotubes}}

\author{Iek-Heng Chu} 
\affiliation{Department of Physics and Quantum Theory Project, University of Florida, Gainesville, Florida 32611, United States}
\author{Dmitri S. Kilin}
\affiliation{Department of Physics and Quantum Theory Project, University of Florida, Gainesville, Florida 32611, United States}
\affiliation{Department of Chemistry, University of South Dakota, Vermillion, South Dakota 57069, United States}
\author{Hai-Ping Cheng}
\email[Corresponding author: Hai-Ping Cheng,  ]{Email: cheng@qtp.ufl.edu}
\affiliation{Department of Physics and Quantum Theory Project, University of Florida, Gainesville, Florida 32611, United States}

\begin{abstract}
{{\bf Abstract:} We have studied the energetics, electronic structure, optical excitation, and electron relaxation of dinitromethane molecules (CH$_{2}$N$_{2}$O$_{4}$) adsorbed on semiconducting carbon nanotubes (CNTs) of chiral index ($n$,0) ($n=7$, 10, 13, 16, 19). Using first-principles density functional theory (DFT) with generalized gradient approximations and van der Waals corrections, we have calculated adsorption energies of dinitropentylpyrene, in which the dinitromethane is linked to the pyrene via an aliphatic chain, on a CNT. A $75.26 \kjmol$ binding energy has been found, which explains why such aliphatic chain-pyrene units can be and have been used in experiments to bind functional molecules to CNTs. The calculated electronic structures show that the dinitromethane introduces a localized state inside the band gap of CNT systems of $n=10$, 13, 16 and 19; such a state can trap an electron when the CNT is photoexcited. We have therefore investigated the dynamics of intra-band relaxations using the reduced density matrix formalism in conjunction with DFT. For pristine CNTs, we have found that the calculated charge relaxation constants agree well with the experimental time scales. Upon adsorption, these constants are modified, but there is not a clear trend for the direction and magnitude of the change. Nevertheless, our calculations predict that electron relaxation in the conduction band is faster than hole relaxation in the valence band for CNTs with and without molecular adsorbates. }
\end{abstract}

\maketitle

\section{Introduction}
Single-walled carbon nanotubes (CNTs) are formed by rolling monolayer graphene sheets in certain directions, 
characterized by the chiral vector $C_{h}=(n,m)\equiv n{\bf a}_{1}+m{\bf a}_{2}$ where ${\bf a}_{1}$ and ${\bf a}_{2}$ are
the lattice vectors for the graphene sheet\cite{CNT_book}. This quasi-one-dimensional material is metallic when $n-m$ is divisible by 3; otherwise it is semiconducting.

In the past two decades, CNTs have received intense attention\cite{optical1,optical2,gas1,gas2,gas3,gas4,gas5,gas6,pvc1,pvc2,pvc3,pvc4,pvc5,cnt1,cnt2} due to their unique tunable electronic properties, which lead to remarkable mechanical, thermal and optical features. A variety of potential applications of CNTs have been studied by many experimental and theoretical groups in recent years, including the CNT-based nanosensing applications such as gas sensors\cite{gas1,gas2,gas3,gas4,gas5,gas6} which can detect gases such as O$_{2}$\cite{gas1}, H$_{2}$\cite{gas2,gas4} or NO$_{2}$\cite{gas5,gas6}; and photovoltaic solar cells\cite{pvc1,pvc2} in which the CNTs serve as electron acceptors
and the conjugated polymers, for example P3OT, as electron donors. More recent experiments also used the Si-CNT junctions to achieve the solar cell with high efficiency\cite{pvc3,pvc4}. 

The interaction of CNTs with atoms or molecules is the key
ingredient for utilizing CNTs as materials with desired functions
and properties. There are two basic types of particle adsorption,
chemisorption and physisorption. In chemisorption, the adsorbate, an atom or a molecule, is covalently bonded to the CNT. It either transfers electrons to or extracts them from the nanotube resulting in an enhanced conductance\cite{K_dope1,K_dope2}. In physisorption, the van der Waals interaction becomes the dominant interaction between the adsorbate and the CNT, and the electronic
structure of the CNTs only changes slightly; however, the dopants can
introduce some localized states that are placed inside the band gap
of semiconducting CNTs\cite{gap_state}. Depending on the position of the impurity state inside the band gap, 
$n$-type or $p$-type semiconducting CNT systems can
be made. For problems related to photophysics and chemistry, these
localized, mid-gap states open additional relaxation pathways to
facilitate the electron transfer upon
photo-excitations\cite{Micha_2010}. The rates for these processes,
which measure how fast the electron in the CNT is transferred to
the chemical species, are important quantities in the dynamics of
such relaxation aided by lattice vibration. Therefore, an
understanding of rates and hence the relaxation processes is of fundamental importance in device applications like
photovoltaic solar cells and organic light-emitting diodes.

There are several treatments that can be used to study electron dynamics of interest. They are (i) complete active space configuration interaction (CAS-CI)\cite{CAS-CI} or time-dependent density functional theory (TDDFT)\cite{TDDFT}, which is combined with dynamics methods such as Born-Oppenheimer molecular dynamics (MD), Car-Parrinello MD\cite{CPMD}, force-field MD etc; (ii) the fewest-switches-surface-hopping in the time-dependent Kohn-Sham approach\cite{NAMD,TDKS1,TDKS2}; and (iii) the reduced density matrix formalism\cite{MayWILEY2000} combined with DFT \cite{Micha_2008,
Micha_2010,HKtheorem,K-S_eqn}. While CAS-CI or TDDFT based dynamics approach can be applied to study excited states and photo-induced dynamics with high accuracy, their high computational costs limit their application to comparatively small-scale systems. The reduced density matrix formalism, which includes weak electron-phonon couplings, is an appropriate theoretical tool to study phonon-assisted processes, as it requires the least computation effort among the approaches mentioned above. Previously, Micha \textit{et al.} has utilized this method to study the relaxation dynamics of Si surfaces with adsorbed Ag clusters\cite{Micha_2010}, and the results are consistent with experiments. Besides, T.M. Inerbaev \textit{et al.} have applied this method to TiO$_{2}$ surfaces\cite{TiO2}, and J. Chen \textit{et al.} have investigated the doped silicon quantum dots using this approach\cite{doped_Si}.

In this paper, we report our investigations using first-principles
methods to study the physisorption of dinitromethane
CH$_{2}$N$_{2}$O$_{4}$ on semiconducting zigzag single-walled CNTs. This molecule contains a nitro group, which is
known as a good electron acceptor. We have calculated the adsorption energies of dinitromethane or dinitropentylpyrene attached to various CNTs, electronic structures before and after molecular adsorption, and also photo-excitation and subsequent phonon-assisted intra-band relaxation dynamics for dinitromethane physisorbed with CNTs. In particular, we have investigated such relaxation processes in the pristine CNTs since they have been studied in recent experiments\cite{relax_1ps,relax_1ps2,CNT_100fs,CNT_100fs_2,CNT_fs}.

The work presented here is organized as follows. The model and the methods 
are introduced in section (II); results are shown in section (III),
which contains three parts: (A) energetics and electronic structures of the systems; (B) \textit{ab-initio} molecular dynamics and non-adiabatic couplings and (C) dynamics of the
phonon-assisted relaxation upon initial photo-excitations; and conclusions are in section (IV).

\begin{figure*}[htpp]
\includegraphics[width=12cm]{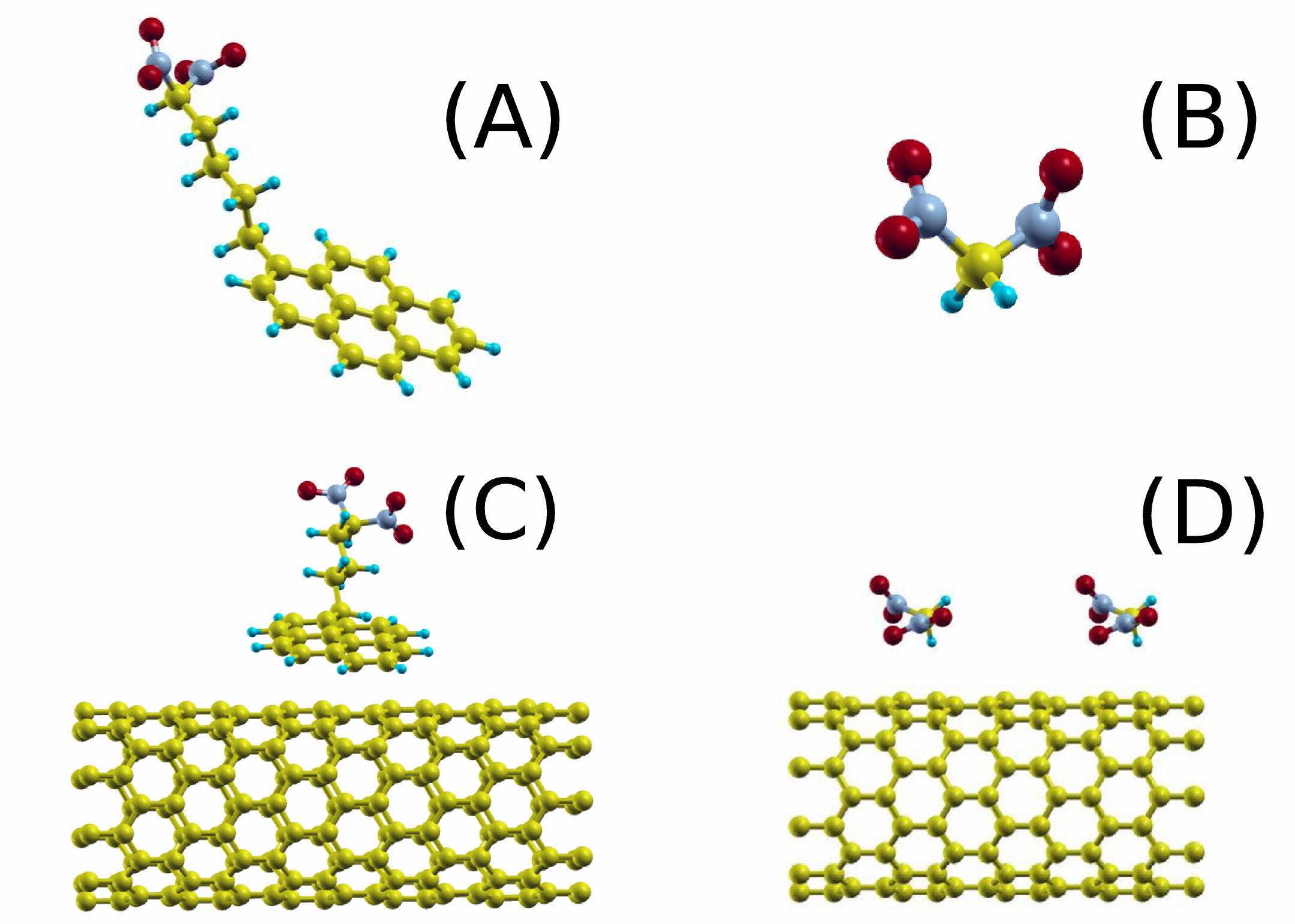}
\caption{Panel ({\bf A}): the dinitropentylpyrene molecule, where the colors red, gray, blue and yellow show
oxygen, nitrogen, hydrogen and carbon atoms, respectively; 
panel ({\bf B}): the dinitromethane molecule; 
panels ({\bf C, D}): CNT with a dinitropentylpyrene and
with a dinitromethane, respectively.
The figure is created using XCrysden\cite{XC}. }\label{cnt_sys}
\end{figure*}

\section{Models, Methods and Computational Details}
We have set up two types of theoretical models. The first one consists of a semiconducting ($n$,0) CNT and a dinitropentylpyrene, in which the dinitromethane CH$_{2}$N$_{2}$O$_{4}$ is linked, via an aliphatic chain, to the pyrene C$_{16}$H$_{10}$ (Fig.\ref{cnt_sys}A) that has been adsorbed on the CNT. The second one consists of the same ($n$,0) CNT but only a dinitromethane (Fig.~\ref{cnt_sys}B) directly adsorbed on
the CNT. Since the sizes of the adsorbate in both cases are different, the corresponding
CNT supercells for adsorbates in the first and the second cases have been constructed to contain five and two CNT
primitive cells, or 200 and 80 carbon atoms, respectively. Such a choice has guaranteed that the interactions between the molecule and its translational images are sufficiently weak. The two systems are shown in Fig.~\ref{cnt_sys}C and Fig.~\ref{cnt_sys}D.
The first model has been used to understand the enhanced binding between the functioning molecule and the CNT via large
contact area between the pyrene fragment and the CNT, as was realized in experiments\cite{pvc2,pyrene2,pyrene3}, and the second model has served as a simplified model system for investigating electron dynamics upon excitation.

The rest of this section details (A) the density functional theory (DFT) and approximations that have been used for the structural optimizations, energetics and the electronic structures; and (B) the reduced density matrix formalism combined with the DFT that has been employed to study the dynamics of the phonon-assisted intra-band relaxations in both the conduction band and the valence band.

\subsection{Density Functional Theory and Calculations}
\subsubsection{{\bf Density Functional Theory}}

The foundations of DFT state that for an $N$-electron system, all the ground-state physical properties can be determined once the ground-state charge
density of the system is known. In this theory, the total energy of the system can be expressed
as\cite{K-S_eqn}
\begin{eqnarray}
E_{tot}=\sum_{i=1}^{N} &&-\frac{\hbar^{2}}{2m}\int
d{\bf r} \,\phi^*_{i}({\bf r}) \,\nabla^{2}\phi_{i}({\bf r})\label{Etot}
+ \frac{1}{2}\int\int d{\bf r}\,d{\bf r}'\,\frac{\rho({\bf r})\rho({\bf r}')}{|{\bf r}-{\bf r}'|}
\\ \nonumber
&&+\int d{\bf r}\,V_{ion}({\bf r})\rho({\bf r})
 +\int d{\bf r}\,\varepsilon_{xc}({\bf r})\rho({\bf r}) \nonumber
\end{eqnarray}
where the charge density is
$\rho({\bf r})=\sum_{a=1}^{N}|\phi_{a}({\bf r})|^{2}$,
$V_{ion}({\bf r})$ is the potential from the ions, and
$\varepsilon_{xc}({\bf r})$ is the exchange-correlation functional of $\rho({\bf r})$. 
There are many types of approximations to this functional, including (i) the local density approximation (LDA), under which $\varepsilon_{xc}$ only depends on $\rho({\bf r})$; (ii) the generalized gradient approximation (GGA), where $\varepsilon_{xc}$ depends on both
$\rho({\bf r})$ and $|\nabla\rho({\bf r})|$; (iii) GGA with inclusion of semi-empirical van der Waals (vdW) interaction; and (iv) the vdW-DF functional, which includes the vdW interaction self-consistently.

Minimizing $E_{tot}$ with respect to $\{\phi_{i}({\bf r})\}$ along with the constraints $\int
d{\bf r}\phi^*_{i}({\bf r}) \phi_{j}({\bf r})=\delta_{i,j}$, one
arrives at the Kohn-Sham equation\cite{K-S_eqn}, which reads:
\begin{eqnarray}
&& \left[ -\frac{\hbar^{2}}{2m}\nabla^{2}+V_{\rm eff}({\bf r}) \right] 
\phi_{i}({\bf r})=\epsilon_{i}\phi_{i}({\bf r}),\\
&&V_{\rm eff}({\bf r})=V_{ion}({\bf r})+\int
d{\bf r}'\frac{\rho({\bf r})}{|{\bf r}-{\bf r}'|}+V_{xc}({\bf r}),
\end{eqnarray}
where $\{\phi_{i}({\bf r})\}$ are known as the Kohn-Sham orbitals,
and $V_{xc}({\bf r})$ is the exchange-correlation potential.

Since $V_{\rm eff}({\bf r})$ also depends on the total charge density, which is determined by the Kohn-Sham orbitals $\{\phi_{i}({\bf r})\}$, the
equation has to be solved self-consistently. The ground-state charge density $\rho({\bf r})$ and the total energy can then be computed according to Eq.~(\ref{Etot}).

In general, when the system is in a nonequilibrium excited state, the charge density is composed as $\rho({\bf r})=\sum_{a,b}\rho_{ab}\,\phi^{*}_{a}({\bf r})\phi_b({\bf r})$ with $\rho_{ab}$ the elements of the reduced density matrix in the basis of the Kohn-Sham orbitals. For a system in the ground state, the matrix elements become $\rho_{ab}=\delta_{a,b}f_{a}$ where $f_{a}=1$ for $1\le a\le N$ and zero otherwise. A brief introduction to the reduced density matrix formalism is given in the sub-section (B).

\subsubsection{{\bf Energetics and Electronic Structures Calculations}}

The energetics, electronic structures and structural optimizations 
have been performed using the Quantum Espresso (QE) package\cite{PWSCF},  
which utilizes the Rappe-Rabe-Kaxiras-Joannopoulos (RRKJ) ultrasoft
pseudopotentials\cite{RRKJ} and a plane-wave basis set. To understand how different
exchange-correlation functionals affect the results, four different energy functionals have been examined: 
(a) The LDA exchange-correlation functional as parametrized by Perdew and Zunger\cite{PZ_LDA}; 
(b) The Perdew-Burke-Ernzerhof (PBE) exchange-correlation functional\cite{PBE_GGA} within the GGA method; 
(c) a semi-empirical dispersion term, as proposed by Grimme\cite{vdW-se} 
in conjunction with the PBE-GGA functional (we call it PBE+vdW);  and 
(d) the vdW-density functional (vdW-DF) proposed by Dion \textit{et al.}\cite{vdW-DF1,vdW-DF2}, which includes the vdW interactions in the calculations self-consistently. The PBE-GGA without corrections does not include the van der Waals (vdW) binding\cite{woods}, which is important in our studies. The energy cutoff for the wave functions has been chosen as
40 Ry to ensure total energy convergence. The structures have been
relaxed until the force on each atom is less than 0.01 eV/\AA. Periodic boundary conditions have been applied along the tube axis, while a vacuum layer of 10 \AA~has been added in the other two directions to avoid nonphysical interactions between adjacent images. 

When necessary, the electronic structure calculations have also been performed using the Vienna Ab-initio Simulation Package (VASP)\cite{VASP1,VASP2} DFT code within the projector augmented wave (PAW) method\cite{PAW}. In such calculations, the LDA exchange-correlation functionals or a new version of the vdW-DF which employs the optB86b exchange functionals\cite{vdW-DF1,vdW-DF2,vdW-DF3} (optB86b vdW-DF) have been used. 

To measure the interaction between the CNTs and adsorbates, we have calculated the binding energy $E_b$, 
which is defined as 
\begin{equation}
E_{b}=E_{CNT}+E_{m}-E_{CNT+m} \label{Ebind}, 
\end{equation}
where $E_{CNT}$ and $E_{m}$ are the total energies of the pristine CNT and the molecule, respectively. The binding is stable if $E_{b}$ is positive.

\subsection{Density Matrix Theory and Calculations}
\subsubsection{{\bf Reduced Density Matrix Formalism}}

In this sub-section, we briefly describe the reduced density
matrix formalism that has been used in our investigations. A
thorough description of the theory can be found in references
[\onlinecite{Micha_2010}], [\onlinecite{MayWILEY2000}] and [\onlinecite{Micha_2008}].
For a closed system that consists of a subsystem of interest and a reservoir,
the total Hamiltonian $H$ can be written as a sum of the
system part $H_{S}$, the reservoir part $H_{R}$ , and the
system-reservoir interaction $H_{S-R}$: $H=H_{S}+H_{R}+H_{S-R}$. The
density operator for the whole system reads
$\hat{\rho}_{tot}(t)=|\Phi(t)\rangle\langle\Phi(t)|$, where $|\Phi(t)\rangle$
is the pure state that represents the closed system. Since we are
only interested in the system part, the reduced density operator
(RDO) for the system part can be defined as:
$\hat{\rho}(t)=\textrm{Tr}_{R}\{\hat{\rho}_{tot}(t)\}=\sum_{R}\langle\phi_{R}(t)|\hat{\rho}_{tot}(t)|\phi_{R}(t)\rangle$;
the trace of the density operator in the reservoir space yields 
the RDO for the system part. In the energy representation where
$H_{S}$ is diagonal and its eigenstates $\{|\phi_{n}\rangle\}$ form
an orthonormal basis set, the Markovian equation of motion for the RDO
can be expressed as 
\begin{eqnarray}
&&\frac{d\rho_{ab}(t)}{dt}=-\frac{i}{\hbar}\{[\hat{H}_{S},\hat{\rho}(t)]\}_{ab}
+\Bigl(\frac{\partial\rho_{ab}}{\partial t}\Bigr)_{diss}\label{qme1}\\
&&\Bigl(\frac{\partial\rho_{ab}}{\partial t}\Bigr)_{diss}=-\sum_{cd}R_{ab,cd}\,\rho_{cd}(t)\label{qme2},
\end{eqnarray}
where
$\rho_{ab}(t)\equiv\langle\phi_{a}|\hat{\rho}(t)|\phi_{b}\rangle$ and $R_{ab,cd}$ are
the elements of the reduced density matrix and of the Redfield tensor, respectively.
The Redfield tensor $R$ describes the interaction between the system and 
the reservoir, and the elements of which can be expressed in terms of the Fourier components of the reservoir time-correlation function\cite{Redfield, MayWILEY2000}. The second term on the right hand side of
Eq.~(\ref{qme1}), or Eq.~(\ref{qme2}), captures the irreversible
dynamics of the energy dissipation from the system to the
reservoir.
It is convenient to work for the Redfield tensor in the interaction picture\cite{MayWILEY2000},
\begin{equation}
\Bigl(\frac{\partial\rho_{ab}^{(I)}}{\partial t}\Bigr)_{diss}
=-\sum_{cd}e^{i(\omega_{ab}-\omega_{cd})\Delta t}R_{ab,cd}\,\rho_{cd}^{(I)}(t)\label{Redfield_int},
\end{equation}
with $\rho_{ab}^{(I)}=\exp(i\omega_{ab}\Delta t)\rho_{ab}(t)$. Here,
$\Delta t$ is the time step, which needs to be larger than the time
scale given by the reservoir time correlation function so that the 
Markov approximation holds. The term $\hbar\omega_{ij}=\epsilon_{i}-\epsilon_{j}$ is the electronic
transition energy. Since the phase factors oscillate rapidly, only terms with $\omega_{ab}-\omega_{cd}=0$
survive after the long time average using a time $\tau\gg\Delta t$. This is known as the secular approximation\cite{MayWILEY2000}, which gives rise to two different cases: (i) $a=b$, $c=d$; 
(ii) $a\neq b$, $a=c$, $b=d$. Therefore, Eq.~(\ref{Redfield_int}) can be separated
into two sets of equations that read,\cite{MayWILEY2000}\\

(i) Population transfer:
\begin{eqnarray}
&&\Bigl(\frac{\partial\rho_{aa}}{\partial t}\Bigr)_{diss}=-\sum_{c}R_{aa,cc}\,\rho_{cc}(t)\label{pop1}\\
&&R_{aa,cc}=\delta_{ac}\sum_{e}k_{ae}-k_{ca}\label{pop2}
\end{eqnarray}

(ii) Coherence dephasing:
\begin{eqnarray}
&&\Bigl(\frac{\partial\rho_{ab}}{\partial t}\Bigr)_{diss}=-R_{ab,ab}\, \rho_{ab}(t)\\
&&R_{ab,ab}=\sum_{c}\frac{1}{2}(k_{ac}+k_{bc})+\gamma_{0},
\end{eqnarray}
where $k_{ab}$ is the transition rate from $|\phi_{a}\rangle$ to
$|\phi_{b}\rangle$ and $\gamma_{0}$ is the pure dephasing constant\cite{dephase}. When the electrons comprise
the system while the ions are considered as the reservoir\cite{exc-ph}, the rate
coefficients $k_{ab}$ are \cite{Micha_2008,Micha_2010},
\begin{equation}
k_{ab}=\frac{1}{\hbar^{2}}|V_{ab}|^{2}J(\omega_{ab})[n(\omega_{ab},T)+1]\label{k_eph},
\end{equation}
where $V_{ab}=-i\hbar\langle\phi_{a}|\partial/\partial t|\phi_{b}\rangle$
is called the non-adiabatic coupling between $|\phi_{a}\rangle$ and
$|\phi_{b}\rangle$, which depends on the ionic trajectories
$\{{\bf R}(t)\}$. The factor $J(\omega)=\sum_{j}\delta(\omega_{j}-\omega)$ is the
vibrational density of states and
$n(\omega,T)=[\exp(\hbar\omega/k_{B}T)-1]^{-1}$ is the phonon
occupation number for a mode with angular frequency $\omega$ at temperature $T$.
Note that three parameters need to be determined beforehand, (\textit{i}) the
vibrational density of states $J$, (\textit{ii}) the phonon occupation $n$ and 
(\textit{iii}) non-adiabatic couplings $V_{ab}$.

\subsubsection{{\bf Dynamics of the Intraband Relaxation}}

We have used Eq.~(\ref{pop1}), along with Eq.~(\ref{pop2}) and
Eq.~(\ref{k_eph}) in the previous sub-section to study the dynamics of the electronic
relaxation in both the conduction band and the valence band upon
initial photo-excitations. The transition dipole moments and
oscillator strengths of the various electronic excitations have been
calculated and used to determine the most optically active
electronic state\cite{Osc-str}. The transition dipole moment between states $a$ and $b$, ${\bf D}_{ab}$, is defined as ${\bf D}_{ab}\equiv\langle\phi_{a}|{\bf r}|\phi_{b}\rangle$, and the oscillator strength is expressed as 
\begin{equation}
f_{ab}=\frac{2m\omega_{ab}|{\bf D}_{ab}|^2}{\hbar}\label{Oscstr} .
\end{equation} 

The non-adiabatic couplings in Eq.~(\ref{k_eph}), which describe the electron-phonon
interactions, have been computed based on an \textit{ab-initio} molecular
dynamics. The system is initially heated up to $300 \K$ by
repeated velocity rescaling, then the dynamics has been performed in the
micro-canonical ensemble, which yields the ionic trajectories. At any
two consecutive time steps $t$ and $t+\Delta t$, we have recalculated the
Kohn-Sham orbitals $|\phi_{a}(t)\rangle$, $|\phi_{b}(t)\rangle$, $|\phi_{a}(t+\Delta
t)\rangle$ and $|\phi_{b}(t+\Delta t)\rangle$ and the couplings are computed using finite difference increment. 
The final non-adiabatic coupling square $|V_{ab}|^{2}$ is approximated as the average over
all those values. The coupling parameter converges if the time window for averaging 
is large enough. Numerical values of the coupling and the convergence of the averaging procedure are reported in the Results section.

After solving Eq.~(\ref{pop1}), we have studied the population distribution in the energy and time domain as well as the charge density distribution in the conduction band as functions of time $t$ and height $z$, which is defined as from the CNT to the C atom in the adsorbate and is perpendicular to the tube axis. The distributions are defined in the following:
A nonequilibrium population distribution in the energy and time domain
reads\cite{Micha_2010}
$n^{(a,b)}(\varepsilon,t)=\sum_{i}\rho^{(a,b)}_{ii}(t)
\delta(\varepsilon_{i}-\varepsilon)$ , where ($a$,$b$) denotes the
initial photo-excitation from state $a$ to state $b$. The change of the
population with respect to the equilibrium distribution is then
expressed as 
\begin{equation}
\Delta
n^{(a,b)}(\epsilon,t)=n^{(a,b)}(\epsilon,t)-n^{eq}(\epsilon).\label{dn}
\end{equation}
This equation describes a population gain when $\Delta n > 0$ and a loss when $\Delta n < 0$ at energy $\epsilon$, 
which corresponds to the electrons and holes.

The charge density distribution in the conduction band as
functions of $z$ and $t$ is defined as\cite{Micha_2010}
\begin{equation}
P_{CB}^{(a,b)}(z,t)=\sum_{i,j \in CB}\rho^{(a,b)}_{i,j}(t)\int
dx\,dy\, \phi^*_{i}({\bf r})\phi_{j}({\bf r}),\label{Pcb}
\end{equation}
where $\phi_{i}({\bf r})$ is the $i^{th}$ Kohn-Sham orbital, and the
indices $i$ and $j$ belong to the conduction band. This distribution reflects the time evolution of the charge transfer 
along the $z$ direction from the CNT into the adsorbate.

Finally, the time evolution of the population of the lowest unoccupied molecular orbital (LUMO) as well as the highest occupied molecular orbital (HOMO) can be fitted to the following equation,
\begin{equation}
P_{e(h)}(t)=1-\exp(-t/\tau^{e(h)})\label{time_eh},
\end{equation}
where $e$ and $h$ refer to the electron in the conduction band and the
hole in the valence band, respectively. The constant $\tau^{e(h)}$ represents the
average relaxation time for the electron (hole), and thus
the dynamics of the electronic relaxation.\\

\section{Results}

\subsection {Energetics and Electronic Structure}
We have prepared two groups of systems that consist of zigzag semiconducting CNTs with chiral indices ($n$,0) ($n=7$, 10, 13, 16, 19) adsorbed with either (1) a dinitropentylpyrene or (2) a dinitromethane, as shown in Fig.~\ref{cnt_sys}.
The ($n$,0) CNTs with $n\geq 10$ have radius large enough such that they be easily made in experiment.  
To obtain ground-state electronic structures and binding energies, we have first optimized structures of these model systems.
For physisorption, it is known that different exchange-correlation functionals may lead to different binding distance between 
the CNT and the adsorbate\cite{woods} and also to different binding energy.
To test this, we have performed calculations using four different functionals 
for the (10,0) CNT with the dinitropentylpyrene that contains a pyrene fragment.
The resulting binding energies computed according to Eq.~(\ref{Ebind}), and the CNT-molecule distances are shown in Table I. 
The binding energy given by the PBE-GGA is only $ 0.59 \kjmol $, and the
equilibrium distance is $ 4.03 \aa$.
When semi-empirical vdW interactions are taken into account\cite{vdW-se} 
along with the PBE exchange-correlation functionals (PBE+vdW), the binding energy
becomes $ 75.26 \kjmol $, and the corresponding distance is $3.23 \aa$. 
When using LDA the binding energy and the distance are $ 36.66 \kjmol $ and $ 3.25 \aa$, respectively.
According to our results, the vdW interaction is important in our system, 
which is the main interaction between the pyrene fragment in the adsorbate and the CNT.
This is consistent with Woods \textit{et al.}'s\cite{woods} results about the interactions between 
benzene molecules and CNTs.
It is clear that the LDA exchange-correlation functional gives results similar to those for the PBE+vdW.
Finally, we have applied to our system the vdW-DF functional \cite{vdW-DF1,vdW-DF2}, 
which includes the vdW interactions in the exchange-correlation functional and thus in the self-consistent procedure. 
The optimized atomic structure and the CNT-molecule distance are found to be 
similar to those obtained from the PBE+vdW or LDA calculations.

\begin{table}[htpp]
\caption{The binding energies $E_{b}$ defined in Eq.~(\ref{Ebind}), the optimized distances $d$
between the (10,0) CNT and the adsorbate dinitropentylpyrene or dinitromethane, using four different exchange-correlation functionals. The values in the parentheses are computed using VASP.}
\begin{tabular}{ c c c c}
\hline \hline
\noalign{\medskip}
$\qquad$ system $\qquad$ & $\qquad$ $ V_{xc} $ $\qquad$ & $\qquad$ $E_{b}$ (kJ/mol) $\qquad$ & $\qquad$ $d$ $(\aa)$ $\qquad$\\
\noalign{\medskip}
\hline
\noalign{\medskip}
  adsorbate       & PBE-GGA  & 0.59  & 4.03 \\
dinitropentyl- & PBE+vdW & 75.26  & 3.23 \\
 pyrene & LDA  & 36.66~(38.50)  & 3.25~(3.14) \\
     & vdW-DF & 84.91~(100.34)   & 3.25~(3.13) \\
  \noalign{\medskip}
 \hline
 \noalign{\medskip}
          & PBE-GGA & 4.61   &  3.96 \\
adsorbate  & PBE+vdW & 26.05  &  3.50 \\
dinitromethane  & LDA & 17.37~(20.84)   &  3.52~(3.36) \\
    & vdW-DF  & 36.37~(34.16)   &  3.67~(3.53) \\
   \noalign{\medskip}
\hline \hline
\end{tabular}
\label{tab1}
\end{table}

In the situation where a single dinitromethane molecule has been physisorbed on the (10,0) CNT, 
we have obtained results that lead to the same trend. 
Without the pyrene fragment, the binding energy now becomes much smaller in cases when 
LDA, PBE+vdW and vdW-DF have been used.
For example, the binding energy in the PBE+vdW case is reduced to only $ 26.05 \kjmol $, 
a factor of three smaller than the value computed in the presence of the pyrene fragment, 
and the optimized distance $d$, defined as the shortest distance between the CNT and the C atom in the dinitromethane molecule, is now $3.50 \aa$.
This suggests that the pyrene fragment enhances the binding between the adsorbate and the CNT, 
and that pyrene is a good choice for anchoring functional groups on CNTs. Note that the binding energy is higher than the thermal energy at $300 \K$, which is about $2.51 \kjmol $. Therefore, we expect that the molecule stay physisorbed around the room temperatures. 

We have also repeated our calculations using VASP.
In these calculations, the LDA exchange-correlation functional and the optB86b vdW-DF, with PAW potentials have been employed.
The binding energies as well as the CNT-adsorbate distances are given in Table I. We have found that the computed distances in all the cases decrease by about 3 to 4\% compared with the LDA values. 

\begin{figure*}[htpp]
\includegraphics[width=14cm]{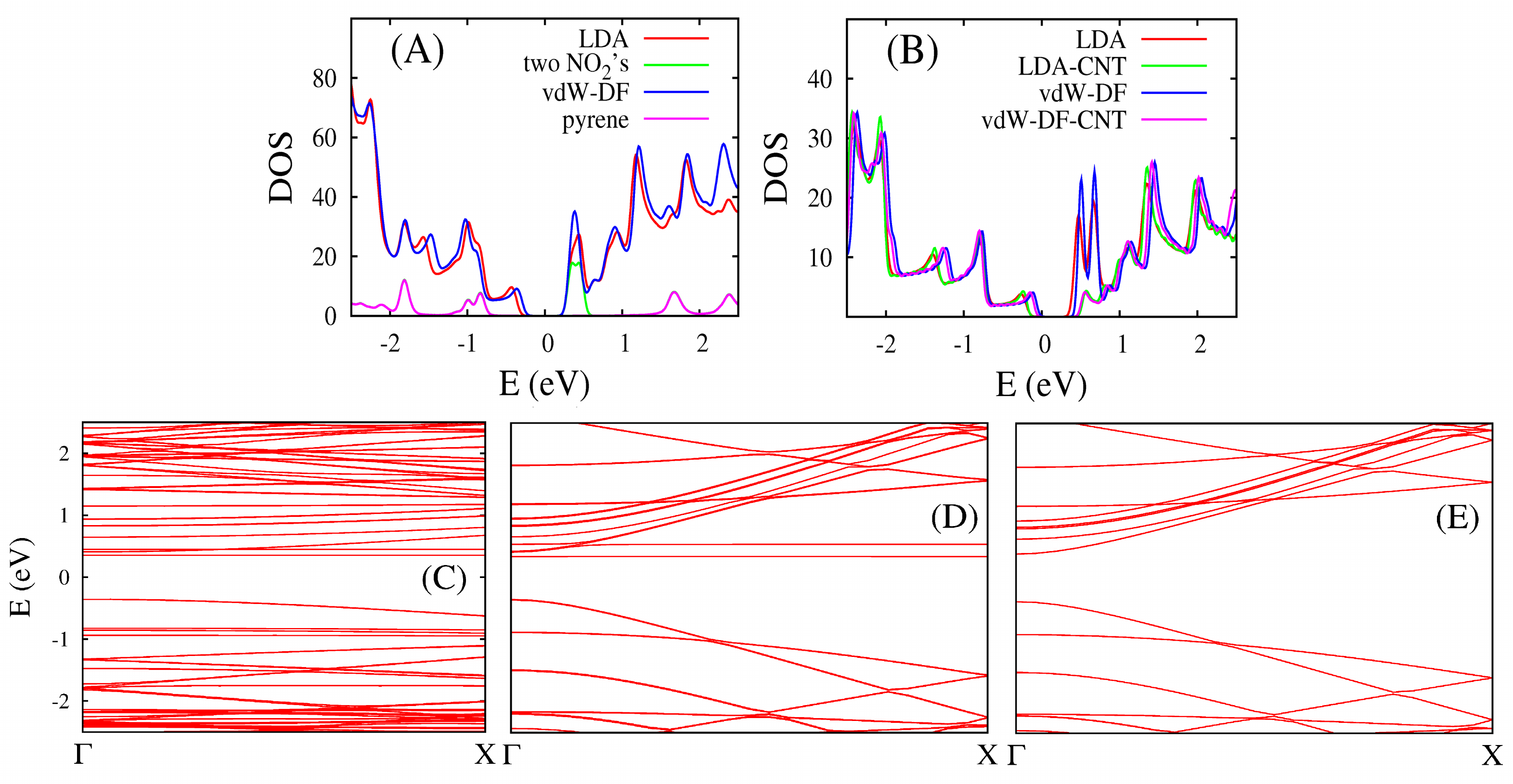}
\caption{ The DOS and band structures in two model systems obtained from LDA and optB86b vdw-DF calculations. 
Panels ({\bf A, C}): The DOS and band structure for (10,0) CNT 
with dinitropentylpyrene. 
In panel ({\bf A}), the projected DOS of the two nitrogen dioxides as well as that of the pyrene fragment are also shown.
Panels ({\bf B, D}): The DOS and band structure for dinitromethane physisorbed on (10,0) CNT.
In panels ({\bf C, D}), the flat bands near the band gap come from the adsorbates.
Panel ({\bf E}): the band structure for the pristine (10,0) CNT for comparison. 
The Fermi energy is set to be zero.}\label{dos_cnt}
\end{figure*}

For each adsorbate depicted in Fig.~\ref{cnt_sys}A or Fig.~\ref{cnt_sys}B, we have computed the band structures and
the density of states (DOS).
Fig.~\ref{dos_cnt} shows the DOS as well as the band structure in each situation from the LDA calculations.
When the pyrene anchoring fragment is present in the adsorbate, the peak immediately above the band gap observed in the DOS 
(Fig.~\ref{dos_cnt}A) is attributed to the two NO$_{2}$ units in the dinitromethane.
Without the pyrene fragment, the dinitromethane is closer to the CNT, 
which leads to an increase in the interaction between the CNT states and the states of the two NO$_{2}$s, 
and thus the peak splits into two (Fig.~\ref{dos_cnt}B).
The corresponding band structures for both situations are shown in Fig.~\ref{dos_cnt}C and Fig.~\ref{dos_cnt}D.
Near the conduction band edge, there are two molecular bands that have vanishing dispersion and do not depend on the $k$-vector. 
One of them is located inside the band gap while the other one is above it.
Here, the important point is that the pyrene fragment in the dinitropentylpyrene does not provide states 
near the band gap, as shown in Fig.~\ref{dos_cnt}A.  
Instead it only acts to anchor the dinitromethane to the (10,0) CNT. The electronic structures obtained from PBE+vdW functionals are similar to the LDA results, and they are not shown here. The DOS plots from the optB86b vdW-DF using VASP are depicted in Fig. 2A and Fig. 2B, for both situations. In the presence of the pyrene fragment, the DOS from the vdW-DF calculations are very similar to the LDA results except that the band gap is now reduced by about 0.15 eV. The same conclusion can be drawn when the dinitromethane molecule is directly physisorbed on the (10, 0) CNT, but the reduction of the band gap is now about 0.2 eV. Based on the calculated electronic structures, we expect that the dynamics of the intra-band relaxation using the adsorbate dinitropentylpyrene or dinitromethane should be very similar. For efficiency, the small-size system that contains the single dinitromethane has been chosen for investigating the dynamics of the phonon-assisted intra-band relaxation upon initial photo-excitation. Also, the following calculations of the dynamics of the intra-band relaxations have been performed using the optB86b vdW-DF.

\subsection{Molecular Dynamics and Non-Adiabatic Couplings}

We have performed the \textit{ab-initio} molecular dynamics for the non-adiabatic couplings in the (10,0) CNT with and without the adsorbate. The time step used in all these calculations is $ 0.5 \fs $. As a first step, we thermalize the systems up to $300 \K$ by repeated velocity rescaling. To guarantee that the corresponding temperature increase is small enough that the thermalization is stable, we have gradually heated systems up to $300 \K$: starting from the ground-state atomic structures, each time we heat systems to $100 \K$ higher than the previous time, with 600 steps between two temperatures.

\begin{figure}
\includegraphics[width=12cm]{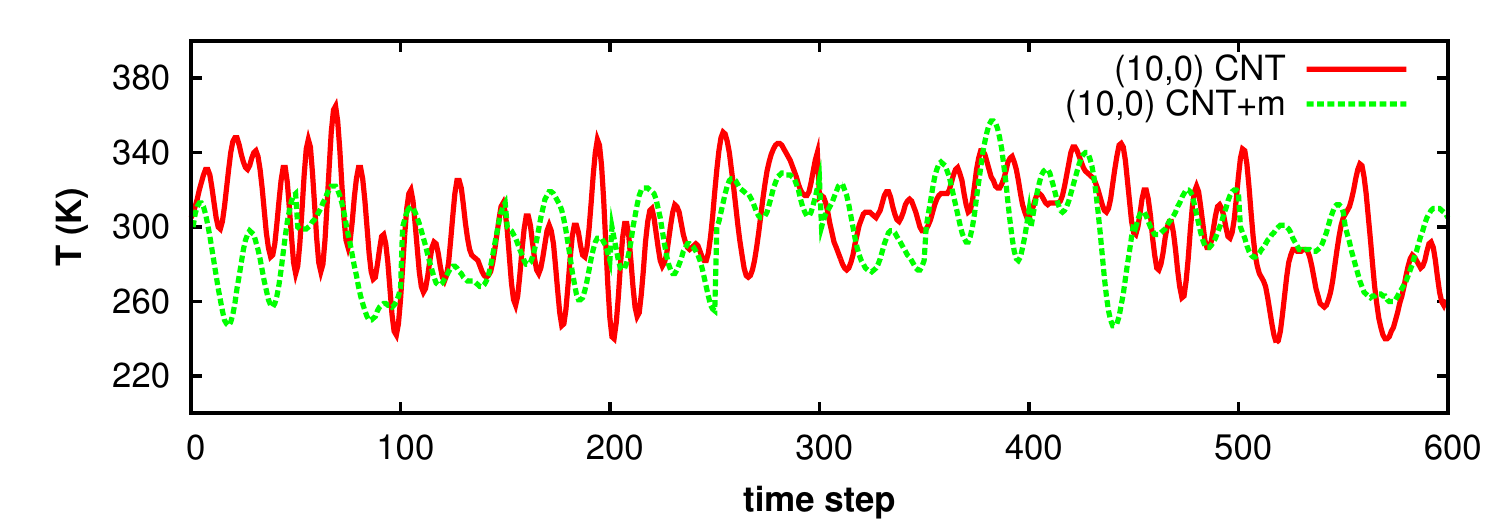}
\caption{The temperature as a function of number of time steps in the \textit{ab-initio} molecular dynamics within micro-canonical ensemble. The red solid line and the green dashed line correspond to the one in (10,0) CNT with and without the molecule, respectively. }\label{QMD_NEV}
\end{figure}

To confirm that systems are thermalized at $300 \K$ and to calculate the non-adiabatic couplings, we then perform 600-step molecular dynamics in the micro-canonical ensemble for both systems. The temperature as a function of number of time steps is plotted for both the pristine and functionalized (10,0) CNT in Fig.~\ref{QMD_NEV}. As shown in the figure, the temperature fluctuates within $20 \%$ around $300 \K$ in both systems.

\begin{figure}
\includegraphics[width=12cm]{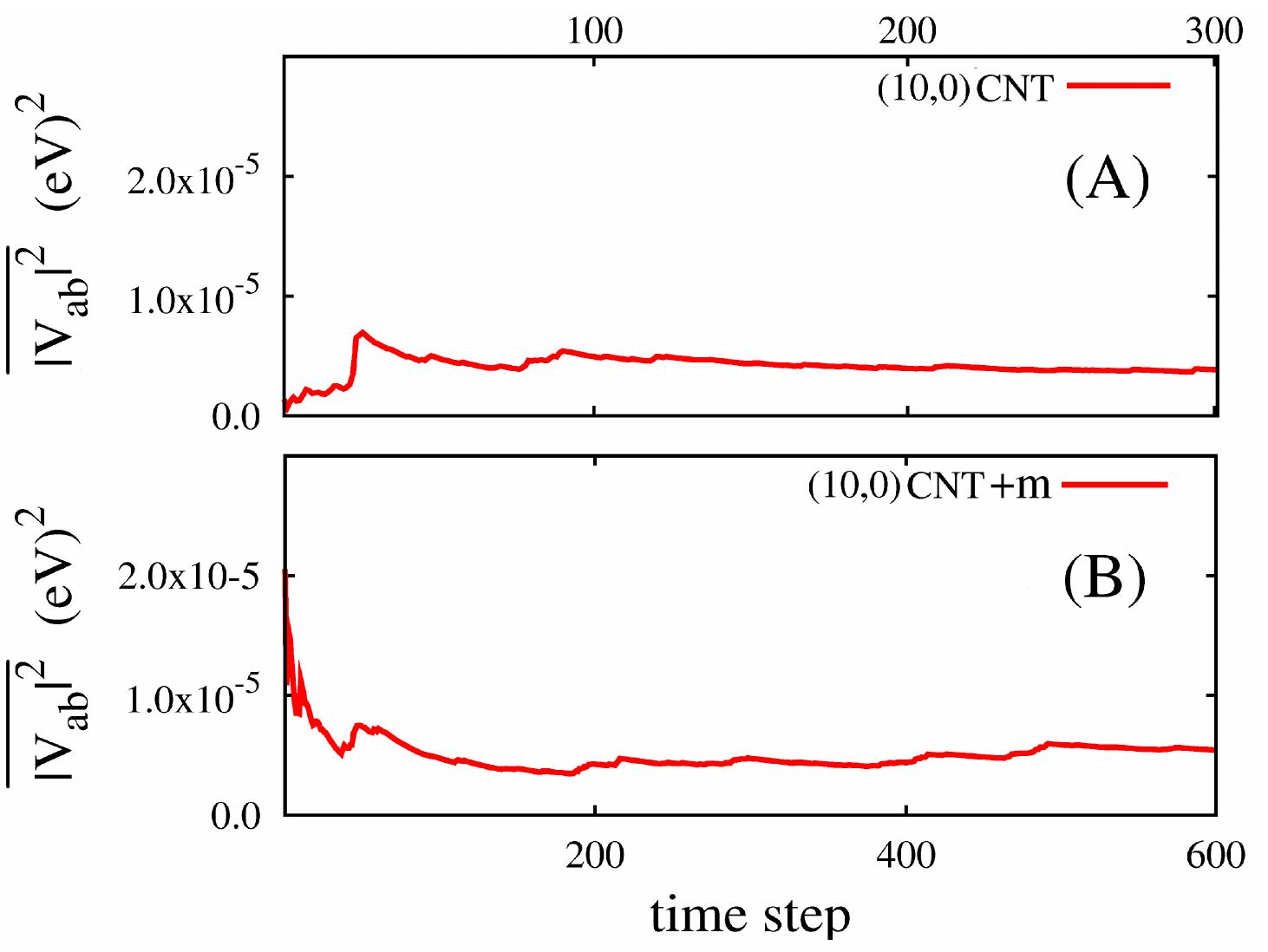}
\caption{The average of non-adiabatic coupling square $\overline{|V_{ab}|^{2}}$ as a function of number of time steps in ({\bf A}) pristine (10,0) CNT and ({\bf B}) (10,0) CNT with the molecule.
Here, $a$ refers to HOMO, and $b$ refers to the HOMO$-3$ and HOMO$-2$ 
for the pristine (10,0) CNT and the (10,0) CNT plus molecule, respectively. }\label{Vab_cnt100}
\end{figure}

Next, at any two consecutive time steps, the non-adiabatic couplings between any two states $a$ and $b$ in the conduction band and/or in the valence band are computed, and the final coupling squares $|V_{ab}|^{2}$ are approximated as the average over all those values. As an example, Fig.~\ref{Vab_cnt100}A shows the convergence of the coupling square of a pristine (10,0) CNT for $a$=HOMO and $b$=HOMO$-3$. The converged value is $ 3.8 \times 10^{-6} (\textrm{eV})^2 $. For comparison, the coupling square in the (10,0) CNT-adsorbate system is given in Fig.~\ref{Vab_cnt100}B. The converged value, $ 5.4 \times 10^{-6} (\textrm{eV})^2 $, is larger than in the pristine (10,0) CNT.

\subsection{Dynamics of the Phonon-Assisted Relaxation}
If the system is optically excited at $t<0$ by steady light with frequency $\Omega=(\epsilon_{b}-\epsilon_{a})/\hbar$,
an electron can be excited from state $a$ in the valence band to state $b$ in the conduction band. For $t\geq 0$, the light is turned off and the electron begins the relaxation towards the LUMO and the hole towards the HOMO.

\begin{figure*}[htpp]
\includegraphics[width=14cm]{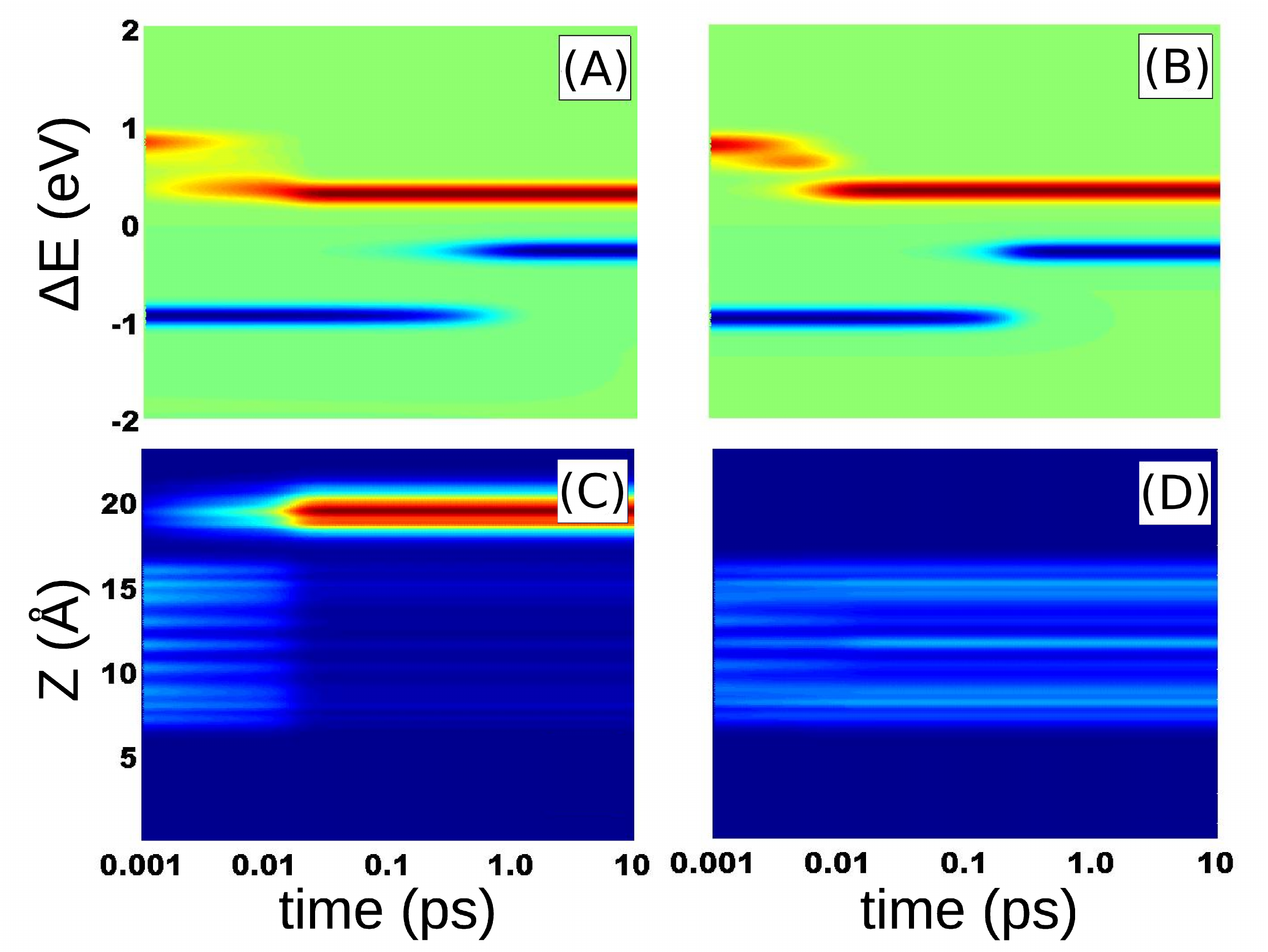}
\caption{Panels ({\bf A, B}): Isocontours of the change of the population
distribution $\Delta n(\Delta\epsilon , t) $ in Eq.~(\ref{dn}) from the dynamics of similar
initial photo-excitations for the (10,0) CNT ({\bf A}) with and ({\bf B}) without dinitromethane, 
in the energy-time domain.
The red, blue and green areas represent gain, loss and no change of population 
relative to the equilibrium distribution respectively.
The initial photo-excitations are from HOMO$-2$ to LUMO$+5$ in ({\bf A}) and from HOMO$-2$ to
LUMO$+3$ in ({\bf B}).
Panels ({\bf C, D}): Spatial distributions of the charge density in conduction band
$P_{CB} (z, t) $ in Eq.~(\ref{Pcb}) for the (10,0) CNT ({\bf C}) with and ({\bf D})
without dinitromethane. 
Color from blue to red indicates the charge density values from 0 to 1,
and the direction $Z$ points from the (10,0) CNT to
the adsorbate.}\label{dynamics}
\end{figure*}

In the pristine (10,0) CNT, the electron at HOMO$-2$ is initially photo-excited to LUMO$+3$, which is the most optically active state according to the oscillator strength values. The intra-band relaxation dynamics is then investigated in the pristine (10,0) CNT. Fig.~\ref{dynamics}B depicts the change of the nonequilibrium population distribution in the energy and time domains. The electron relaxation in the conduction band takes place rapidly. The LUMO starts to gain the electron population at the time of $1 \fs $ and arrives its maximum at about $t=10 \fs $. However, the hole population is not completely transferred to the HOMO until about $t=1 \ps $, which is much slower than the electron population transfer. Fig.~\ref{dynamics}D describes the time evolution of the charge density distribution in the conduction band along the $z$ direction. The change of such distribution completes at about $t=10 \fs $.

\begin{table*}[htpp]
\caption{The electron (hole) relaxation time constants $\tau^{e(h)}$ defined in Eq.~(\ref{time_eh}) for different pristine ($n$,0) CNTs, with given initial and final electron-hole pair excitation transition energies $\Delta E_i$ and $\Delta E_f$, respectively; $f_{if}$ defined in Eq.~(\ref{Oscstr}) is the oscillator strength corresponding to the initial photo-excitation.}
\begin{tabular}{ c  c  c  c  c  r  r }
 \hline \hline
 \noalign{\medskip}
 $n$ & transition & $\quad$ $f_{if}$ $\quad$ & $\Delta E_{i}$(eV) & $\Delta E_{f}$(eV) & $\tau^{e}$(fs) &$\tau^{h}$(fs)  \\
 \noalign{\medskip}
\hline
\noalign{\medskip}
 \ 7 \ &  HO$\,-5\,$ $\rightarrow$ LU+6 & 0.25 & 3.71 &  0.70 & 121 & 595 \\
 \ 10 \ & HO$\,-2\,$ $\rightarrow$ LU+3 & 0.82 & 1.76 &  0.65 &  12 & 416 \\ 
 \ 13 \ & HO$\,-6\,$ $\rightarrow$ LU+2 & 0.99 & 2.69 &  0.53 &   1 &  65 \\
 \ 16 \ & HO$-12$ $\rightarrow$ LU+3 & 1.50 & 2.95 & 0.45 &   4 & 199 \\
 \ 19 \ & HO$-13$ $\rightarrow$ LU+2 & 1.73 & 2.67 & 0.37 &  10 &  81 \\
 \noalign{\medskip}
\hline \hline
\end{tabular}
\label{tab2}
\end{table*}

The average electron and hole relaxation time constants for pristine (10,0) CNT obtained using Eq.~(\ref{time_eh}) 
are given in Table II for initial photo-excitations selected according to oscillator strength values.
The relaxation time constant for the electron is about 35 times smaller than for the hole.
We have also repeated our calculations in the pristine (7,0), (13,0), (16,0) and (19,0) CNTs, 
with results also shown in Table II. There are a few things worth pointing out: First, the relaxation times for the electrons are smaller than those of the holes in all the CNTs; an electron in the conduction band relaxes more rapidly than the hole in the valence band, which agrees with Habenicht \textit{et al.}'s results\cite{TDKS2} of intra-band relaxations for (7,0) CNT. In their paper, they claim that the hole in the valence band interacts more strongly with the radial-breathing phonon mode (RBPM) than with the G-type longitudinal optical phonon mode (LOPM) in the CNT, while the electron in the conduction band mainly interacts with the latter mode. The transition rate between two states depends on the non-adiabatic coupling as well as the number of phonons at the state energy difference. In our studies, the corresponding phonon energy for the LOPM is about $180 \meV $ while it is less than $80 \meV $ for the RBPM, in all the pristine CNT systems. We have found that the energy differences for adjacent, non-degenerate states $\epsilon_{i+1}-\epsilon_{i}$ are closer to the LOPM phonon energy, which suggests that the LOPM plays an important role in the phonon-assisted relaxations. In the (10,0), (13,0) and (16,0) CNTs, we have also found that the non-adiabatic couplings for adjacent conduction band states are about an order of magnitude larger than those for valence band states. In (7,0) and (19,0) CNT, the couplings for conduction band states are about two times larger than those for valence band states. Therefore, the electron couples with the phonon stronger than the hole does, which results in a faster electron relaxation dynamics. 

Second, the hole relaxation time constant in the pristine (7,0) CNT is $595 \fs $, 
while the electron relaxation time constant is only $ 121 \fs $.
These are comparable with the time scales computed in reference[\onlinecite{TDKS2}].
Third, the calculated hole relaxation time constants vary between $ 65 \fs $ to $ 595 \fs $.
The relaxation time constants of the electron, on the other hand, change from $ 1\fs $ to about $ 121 \fs $. Overall, the relaxation time constants calculated in the pristine CNTs agree well with the experimental time scales\cite{relax_1ps,relax_1ps2,CNT_100fs,CNT_100fs_2,CNT_fs}, which vary from less than $ 100 \fs $ to about $ 1 \ps $ 
due to the complicated local environments in the samples. 

\begin{figure}[htpp]
\includegraphics[width=7.5cm]{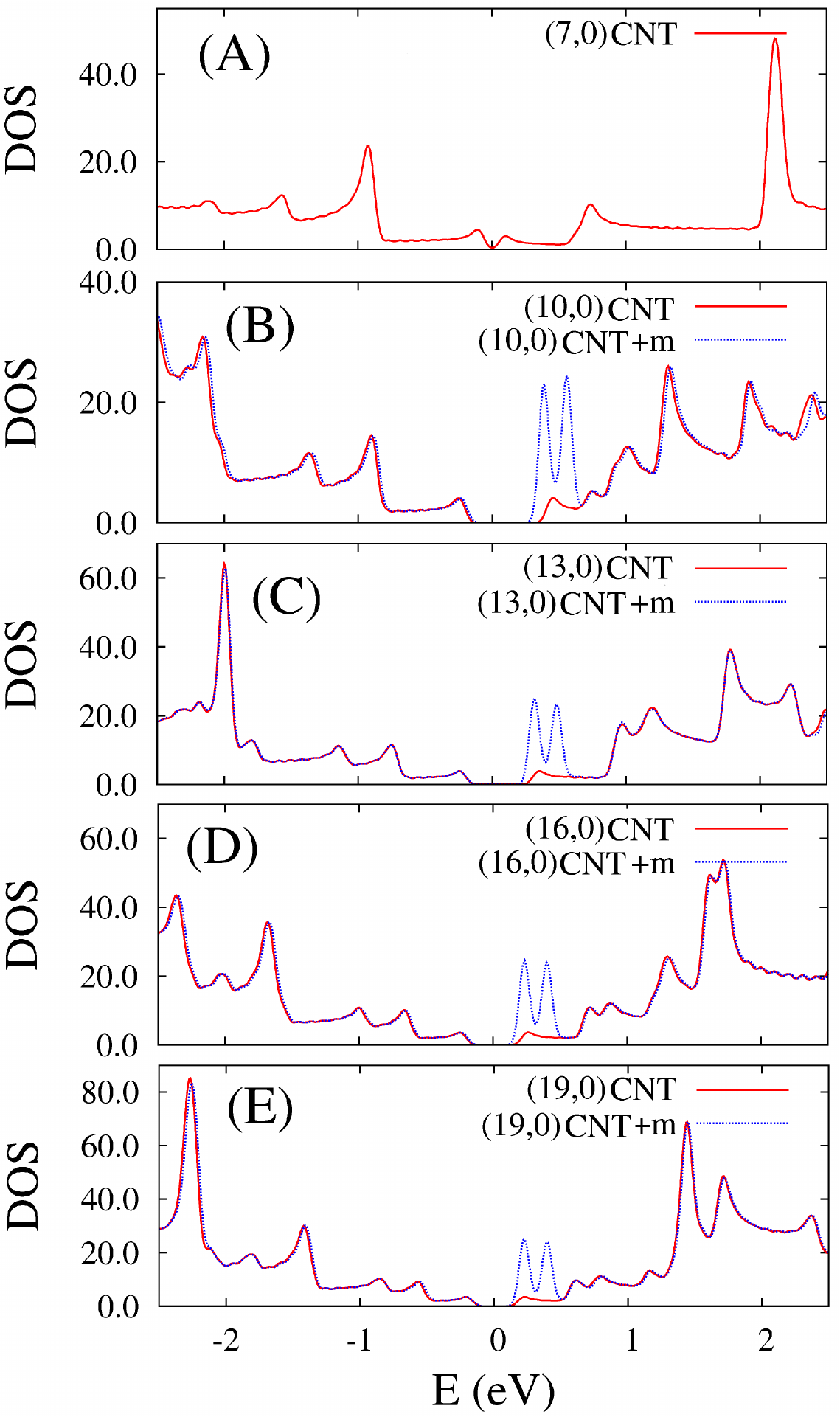}
\caption{The DOS for different pristine ($n$,0) CNTs with ({\bf A}) $n=7$, ({\bf B}) $n=10$, ({\bf C}) $n=13$, ({\bf D}) $n=16$ and ({\bf E}) $n=19$. The DOS for ($n$,0) CNTs physisorbed with the dinitromethane are also given in panels ({\bf B})--({\bf E}).}\label{dos_cnts}
\end{figure}

\begin{table*}[htpp]
\caption{The electron (hole) relaxation time constants
$\tau^{e(h)}$ for different ($n$,0) CNTs
upon adsorption of a dinitromethane molecule, 
with similar initial photo-excitations as those
shown in Table II for pure CNTs.\\}
\begin{tabular}{ c  c  c  c  c  r  r  }
 \hline \hline
 \noalign{\medskip}
 $n$ & transition & $\quad$ $f_{if}$ $\quad$ &$\Delta E_{i}$(eV) & $\Delta E_{f}$(eV) & $\tau^{e}$(fs) & $\tau^{h}$(fs) \\
\noalign{\medskip}
\hline
\noalign{\medskip}
10  & HO$\,-2\,$ $\rightarrow$ LU+5 & 0.53 & 1.76 & 0.58 &   6  & 622 \\
 13 & HO$\,-7\,$ $\rightarrow$ LU+4 & 1.19 & 2.69 & 0.50 &   2  & 123 \\
 16 & HO$-12$ $\rightarrow$ LU+5 & 1.33 & 2.94 & 0.42 &  6  & 107 \\
 19 & HO$-13$ $\rightarrow$ LU+4 & 1.84 & 2.68 & 0.37 & 11  & 169 \\
\noalign{\medskip}
\hline \hline
\end{tabular}
\label{tab3}
\end{table*}

\begin{figure}
\includegraphics[width=10cm]{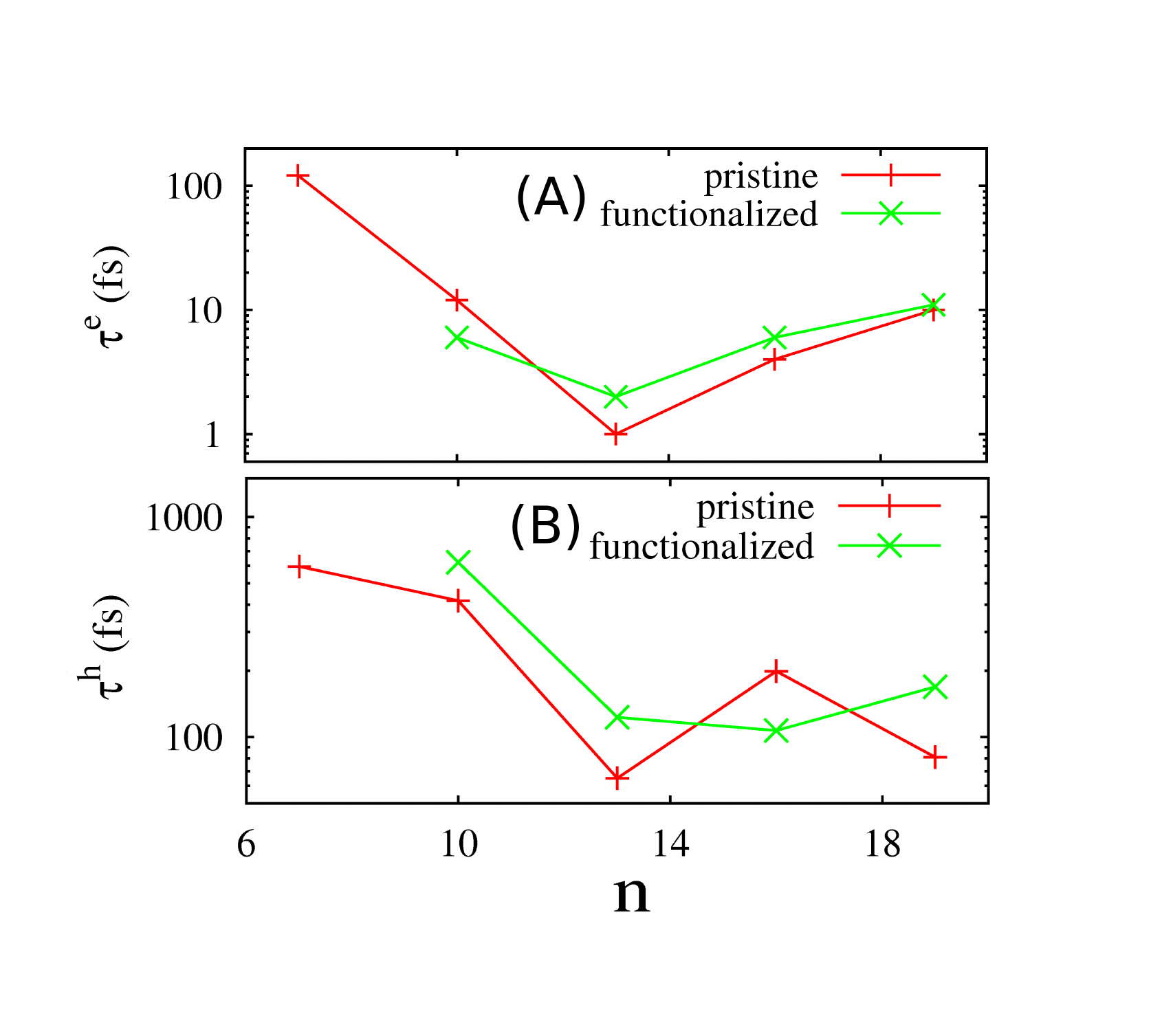}
\caption{Panel ({\bf A}): The electron relaxation time constant $\tau^e$, and ({\bf B}): The hole relaxation time constant $\tau^h$, versus the chiral index $n$, for the pristine ($n$,0) CNTs and for ($n$,0) CNT functionalized by dinitromethane.}\label{rates.n}
\end{figure}

With the single dinitromethane physisorbed on the ($n$,0) CNTs, where $n=10$, 13, 16 and 19, we have also calculated the electron and hole time constants upon initial photo-excitations as in the pristine ($n$,0) CNT cases. Once again, the corresponding oscillator strength value in each case is large, indicating that those excitations are optically active. The results are given in Table III. Upon adsorption, our studies of electronic structures show that there is a gap state provided by the adsorbate in each case, as shown in Fig.~\ref{dos_cnts}, and the electron relaxation time constants are much smaller than the hole relaxation time constants. Again, this is due to larger non-adiabatic couplings in the conduction band than in the valence band. The relaxation time constants as functions of the chiral index $n$, which in this situation is proportional to the tube radius, are visualized in Fig.~\ref{rates.n}. It is clear that the relaxation time constants depend on the tube radius. Compared with the results of the intra-band relaxations in pristine ($n$,0) CNTs, we have found that the electron and hole relaxation time constants are modified, but there is not a clear trend for the change. 

The electron relaxation in all the cases upon adsorption occurs very rapidly. Typically this process completes at less than $ 10 \fs $. When $n=10$, the presence of the gap state facilitates the electron relaxation by a factor of two. When $n=13$, however, the electron relaxation time does increase by a factor of two. When $n=16$ and 19, the electron relaxation times are comparable with those in the corresponding pristine CNT cases. The hole relaxations, on the other hand, are all on sub-picosecond time scales. Except for $n=10$ where the hole relaxation time is comparable with the pristine CNT value, all the hole relaxation times differ by a factor of two when they are compared with the values in pristine CNT cases. In particular, the hole relaxation time constant in both $n=13$ and 19 upon adsorption is doubled, while it is reduced by half when $n=16$. 

Since the LUMO is the gap state in all the four cases and the initially photo-excited electron is in the CNT, when the intra-band relaxation in the conduction band is complete, there is an electron transfer from the CNT to the adsorbate. As an example, Fig.~\ref{dynamics}C shows such process of the photo-induced charge transfer in the (10,0)  CNT-adsorbate system. It is clear that the photo-excited electron originally in the CNT eventually hops to the adsorbate, leaving a hole inside the CNT. In each case, the electron transfer occurs at the same time scales as the intra-band relaxation in the conduction band, i.e. $\tau^{e}$s. This process takes about $10 \fs $ or less in all the cases.

Note that in DFT calculations there is an issue about the band gap error, which is important in the studies of inter-band relaxation where the photo-excited electron in the conduction band recombines with the hole in the valence band.
In the semiconducting CNTs, such relaxation is on time scales of $10 \ps $ or even  $100 \ps $\cite{relax_1ps,interband}, 
which is at least an order of magnitude larger than the intra-band relaxation aided by phonons.
In the present work, since we mainly study the dynamics of the latter, we do not expect that this issue significantly changes our results. However, more accurate calculations such as the hybrid functional calculations\cite{HSE} or the GW method\cite{GW_Hedin,GW_Louie} to correct the band gap will be applied in the future.
In addition, We assume that upon photo-excitations, the geometrical distortion is small so that the Huang-Rhys factors\cite{huang}, which measure the contributions of phonon modes to the reorganization energy, are negligible. We also assume that the electron-phonon interaction is weak such that the intra-band relaxation dynamics can be studied in the Redfield theory within the Markov approximation. Such approximations are justified by the fact that the calculated time constants for pristine CNTs are in good agreement with experiments. However, more rigorous approach such as the nonequilibrium Green's function based technique, where those interactions are described by the non-Markovian relaxation terms in Kadanoff-Baym equations\cite{NMKBE}, can be applied in the future. Finally, in our calculations we have neglected excitonic effects concerning the interaction between the photo-excited electron and the hole. With all these approximations mentioned above, it is surprisingly good that our results agree well with the experiments for the pristine semiconducting ($n$,0) CNTs\cite{relax_1ps,CNT_100fs,CNT_100fs_2}. However, more advanced studies including these effects will be addressed in the future.

\section{Conclusions}
We have studied semiconducting ($n$,0) CNTs for $n=10$, 13, 16 and 19 that are functionalized by a physisorbed electron-accepting functional group, dinitromethane or CH$_2$N$_2$O$_4$. We find that the functional group can be firmly attached to the tube using a pyrene anchoring fragment.  The functional group contributes additional states in the band gap of the CNT, 
but the pyrene fragment acts only as an anchor that enhances the binding between the dinitromethane and the CNT. 
In order to reduce computational intensity, the set of smaller systems, CNTs with the functional group only, has been used for further investigations. We find that an excited state of the functionalized CNT undergoes electronic dynamics coupled to thermal lattice vibrations, and electronic energy dissipates into lattice vibrations. 

Furthermore, our studies of the dynamics of intra-band relaxations indicate that upon initial photo-excitation, an electron from the CNT ends up being transferred to the adsorbate, while a hole stays in the CNT. Electronic-state dynamics demonstrates two effects: (i) electrons and holes lose their energy into thermal vibrations and (ii) electrons and holes migrate in space. Such migration often leads to the formation of a charge separation state. We have quantified electronic relaxation with rate constants. The photo-induced charge transfer completes in about $10 \fs $ in all cases. For pristine CNTs, the calculated intra-band relaxation time constants agree well with the experimental time scales. Our calculations predict that electron relaxation in the conduction band is faster than hole relaxation in the valence band in such CNT systems, with and without the adsorbed dinitromethane molecules. This is due to the stronger electron-phonon interaction in the conduction band than the hole-phonon interaction in the valence band.

Our results have twofold importance: (i) we demonstrate the applicability of the density matrix formulation in the Kohn-Sham DFT framework to a broader range of systems; and (ii) our results provide an important step toward developing solar energy harvesting materials.

\section{Acknowledgements}
I.H. Chu and H.P. Cheng acknowledge support by the Department of Energy (DOE), Office of Basic Energy Sciences (BES) under Contract No. DE-FG02-02ER45995. D.S. Kilin acknowledges support by the DOE, BES-Chemical Sciences under Contract No. DE-AC02-05CH11231, Allocation Award 86185 and US National Science Foundation award EPS0903804. The calculations are performed at NERSC and UF-HPC Center.

\clearpage

\end{document}